\documentclass[usenatbib]{mn2e}

\usepackage[dvips]{epsfig,color}

\usepackage{amssymb,amsmath}

\newcommand{\eg}{{\it e.g.,}}
\newcommand{\ie}{{\it i.e.,}}
\newcommand{\etal}{{\it et al.}}

\newcommand{\ignore}[1]{\relax}

\DeclareMathAlphabet{\mathsfsl}{OT1}{cmss}{m}{sl}
\DeclareMathOperator{\sech}{sech}

\newcommand{\mt}{\ensuremath{M_{\rm total}}}
\newcommand{\fst}{\ensuremath{\tilde{f}}}

\newcommand{\dif}{\mathrm{d}}

\begin{document}

\title{Predicting Steady States of One-dimensional Collisionless
Gravitating Systems}
\author[Ragan \& Barnes]{
Robert J. Ragan\thanks{email:rragan@uwlax.edu}, 
Eric I. Barnes\thanks{email:barnes.eric@uwlax.edu} \\
Department of Physics, University of Wisconsin --- La
Crosse, La Crosse, WI 54601}

\maketitle

\begin{abstract}

Building on the development of a Hermite-Legendre analysis of
one-dimensional gravitating collisionless systems, we present a
technique for determining the steady states of such systems.  This
provides an important component for understanding the physics involved
in the relaxation of these kinds of systems.  As the dark matter
structures in the universe should have traits in common with these
systems, insight into this relaxation can provide clues to larger
astrophysical questions.  For large perturbation strengths, we
determine physically motivated parameter ranges for the simplest
families of steady states as well as their stability.  We also
demonstrate that any set of initial conditions in the linear regime
can be resolved into unique time-independent and time-dependent modes.
Combinations of time-independent modes then describe the steady state of 
any system linearly perturbed from equilibrium.  These results
highlight the importance of initial conditions over relaxation
mechanisms in the evolution of these systems.

\end{abstract}


\section{Introduction}\label{intro}

The current paradigm surrounding the formation of large-scale
structure in the universe relies on the behavior of collisionless dark
matter \citep[\eg][]{setal03,setal05}.  Investigations of
three-dimensional systems, such as individual galactic-scale dark
matter haloes, involve a wide range of evolutionary processes that
contribute to the relaxation from initial conditions to a final
equilibrium state \citep*[\eg][]{nfw96,m98}.  The radial orbit
instability \citep{ma85} along with evaporation and ejection
\citep{bt87} are commonly discussed examples of these processes.  The
sheer variety of processes occurring during the relaxation of a
three-dimensional object significantly complicates any attempt to
disentangle essential behaviors.  Our overall goal is to illuminate
the roles of phase mixing and violent relaxation in self-gravitating
collisionless evolution.

To do this, we concern ourselves only with a one-dimensional version
of a self-gravitating collisionless system.  Dropping to a
one-dimensional system has several advantages.  Immediately, the
evolution of the fine-grained distribution function $f$, which
describes how many particles exist in infinitesimal volumes of phase
space, is completely defined by the following collisionless Boltzmann
(or Vlasov) equation,
\begin{equation}\label{cbe}
\frac{\dif f}{\dif t} = \frac{\partial f}{\partial t} +
v\frac{\partial f}{\partial x} + a(x) \frac{\partial f}{\partial v}=0,
\end{equation}
where $a(x)$ is the acceleration.  This acceleration is a battle
between mass to the left and mass to the right of any location,
\begin{equation}\label{reala}
a(x)=-g\int_{-\infty}^x \lambda(s) \; \dif s + g\int_x^{\infty}
\lambda(s) \; \dif s,
\end{equation}
where $g$ is the gravitational coupling constant and $\lambda$ is the
density distribution.  For a system of mass $M$,
\begin{equation}\label{reall}
\lambda(x)=M\int_{-\infty}^{\infty} f(x,v) \; \dif v.
\end{equation}

The relative simplicity of this version of the Boltzmann equation
provides a manageable starting point for analytical treatment.  A
two-dimensional phase-space structure is straightforward to visualize,
and removes the need for surfaces of section or other techniques for
analyzing higher dimensional spaces.  At the same time,
Section~\ref{theq} presents the details of a one-dimensional,
collisionless equilibrium with a separable form.  Additionally, the
simplicity of the phase space for these systems allows one to take
advantage of highly accurate and efficient $N$-body simulation
schemes.  Unlike three-dimensional situations where issues such as
softening lengths and potential-calculation parameters can blur
insight regarding relaxation processes, simulations of one-dimensional
systems have no free parameters and rely on simple kinematics with
constant acceleration to evolve.  The simplicity of the analysis of
one-dimensional systems has led to decades of work.  \citet{br14} has
a brief discussion of these investigations.  Attacks on systems far
from equilibrium \citep[\eg][]{jw11} and near equilibrium
\citep[\eg][]{rm87} have also been undertaken.  Decompositions using
action-angle variables \citep*{w91,boy11} have also led to insights
into the dynamics of these types of systems.

While the previously highlighted differences are positive for this
work, we freely admit that what follows will be necessarily
unrealistic.  We work under the assumption that the generic
characteristics of phase mixing and violent relaxation are independent
of the dimensionality of the system.  For the results presented here
to have any usefulness in a wider context, these processes must be
simply linked to the collisionless self-gravitating natures of the
systems.  Another shortcoming of this work is that predictive
capabilities are confined to relatively small perturbations from
equilibria.  Initial conditions such as those that would more closely
resemble cosmological conditions evolve non-linearly and lead to
families of time-independent solutions.  Without linearity, such
initial states cannot be uniquely decomposed into these solutions, nor
is it possible to exclude certain families based on initial
conditions.

With these qualifications in mind, what follows is a discussion of a
method for finding time-independent solutions to Equation~\ref{cbe}.
The remainder of this introduction is devoted to reviewing the basics
of the Hermite-Legendre expansion that underlies our analysis.

\subsection{Separable Solution Equilibrium}\label{theq}

Based on the structure of Equation~\ref{cbe}, it is well known
that any function of the specific energy,
\begin{equation}\label{spece}
\epsilon = \frac{v^2}{2} + \phi(x),
\end{equation}
is a solution.  We are specifically interested in the separable
solution to Equation~\ref{cbe}, which is commonly written as,
\begin{equation}\label{ftherm0}
f_0(x,v) = A \sech^2{(\frac{\beta g \mt}{2} x)} e^{-\frac{\beta
v^2}{2}},
\end{equation}
where $\beta$ is an inverse energy, $g$ is the gravitational coupling
constant, $\mt$ is the total mass of the system, and $A =
(g\mt/4)\sqrt{\beta^3/2\pi}$ is the normalization constant.  The
energy scale $\beta$ is related to the total energy of the
equilibrium, $E_0 = 3\mt/2\beta$.  For brevity, we will refer to this
solution as the separable equilibrium from here on.  With this
distribution function, it can be shown that the corresponding
potential is,
\begin{equation}
\phi_0(x)=\ln{(2 \cosh{\frac{\beta g \mt}{2} x})}.
\end{equation}
The importance of this fact lies in its ability to transform
Equation~\ref{ftherm0} into the Boltzmann distribution function,
\begin{equation}
f_0(\epsilon) = A e^{-\beta \epsilon}.
\end{equation}
The Boltzmann nature of the one-dimensional self-gravitating
equilibrium is a vital difference from the three-dimensional case.
This simple form is key to the mathematical approach for dealing
with perturbations to this equilibrium.  Additionally, this form
guarantees that the kinetic temperature of such an equilibrium is
uniform and allows one to view $\beta$ as analogous to $kT$ in a
collisional system's distribution function.

For simplicity, we transform to dimensionless coordinates using the
definitions,
\begin{displaymath}
\chi = \frac{\beta g \mt}{2} x \; , \; 
\varpi = \sqrt{\frac{\beta}{2}} v \; , \; \mbox{and} \; \tau =
\sqrt{\frac{\beta}{2}} g \mt t.
\end{displaymath}
This leaves us to write the scaled equilibrium distribution function as,
\begin{equation}
\fst_0(\chi,\varpi) = \frac{2}{g \mt}\sqrt{\frac{\beta^3}{2}} f_0 =
\frac{1}{2\sqrt{\pi}} \sech^2{\chi} e^{-\varpi^2}.
\end{equation}
Equation~\ref{cbe} transforms to,
\begin{equation}\label{cbe2}
\frac{\partial \fst}{\partial \tau} +
\varpi\frac{\partial \fst}{\partial \chi} + \alpha(\chi) 
\frac{\partial \fst}{\partial \varpi}=0,
\end{equation}
where $\alpha(\chi)$ is the dimensionless acceleration function.  From
Equations~\ref{reala} and \ref{reall}, this acceleration is given by
\begin{equation}\label{dima}
\alpha(\chi)=-\int^{\chi}_{-\infty} \Lambda(\chi^{\prime}) \, \dif
\chi^{\prime} + \int^{\infty}_{\chi} \Lambda(\chi^{\prime}) \, \dif
\chi^{\prime},
\end{equation}
where
\begin{displaymath}
\Lambda(\chi)=\int^{\infty}_{-\infty} \fst(\chi,\varpi) \, \dif
\varpi
\end{displaymath}
is the dimensionless density.

\subsection{Linear Perturbations}\label{lpert}

Our goal is to investigate the relaxation of
systems initially not in equilibrium.  A useful first step in this
direction is to deal with linear perturbations to equilibrium,
\begin{equation}
\fst=\fst_0 + \Delta \fst_1,
\end{equation}
where $\fst_1$ is the perturbing function and $\Delta$ controls the
perturbation strength.  For linear perturbations, we will consider
$\Delta \ll 1$.  The remainder of this section is a brief review
based on work in \citet{br14}.

Using this perturbed $\fst$ in Equation~\ref{cbe} produces a modified
Boltzmann equation for the perturbing function (in terms of the
previously defined dimensionless quantities),
\begin{equation}\label{pcbe}
\frac{\partial \fst_1}{\partial \tau} + \varpi \frac{\partial
\fst_1}{\partial \chi} + \alpha_0(\chi) \frac{\partial \fst_1}{\partial
\varpi} + \alpha_1(\chi) \frac{\partial \fst_0}{\partial \varpi}=0.
\end{equation}
Using Equation~\ref{dima}, it is straightforward to find that,
\begin{displaymath}
\alpha_0(\chi)=- \int_{-\chi}^{\chi}
\int_{-\infty}^{\infty} \fst_0(\chi,\varpi) \, \dif \varpi \, \dif
\chi^{\prime} = -\tanh \chi.
\end{displaymath}
It is also useful to note that,
\begin{displaymath}
\frac{\partial \fst_0}{\partial \varpi} = -2\varpi \fst_0.
\end{displaymath}

We continue by expressing the perturbing distribution function in
terms of Hermite and Legendre functions,
\begin{equation}\label{hlex}
\fst_1=\sum_{m,n}  c_{m,n} \mu \nu
H_m(\varpi) P_n(\tanh \chi) \sech^2 \chi e^{-\varpi^2},
\end{equation}
where $\mu=1/\sqrt{2^m \sqrt{\pi} m!}$ and $\nu=\sqrt{(2n+1)/2}$ are
related to the normalization functions of Hermite and Legendre
functions, respectively.  Also, note that $c_{0,0}=0$ since the
equilibrium has already been removed.  With this identification, the
perturbing acceleration can be written as
\begin{eqnarray}
\alpha_1(\chi) & = &-\int^{\chi}_{-\infty} \Lambda_1(\chi^{\prime}) \, \dif
\chi^{\prime} + \int^{\infty}_{\chi} \Lambda_1(\chi^{\prime}) \, \dif
\chi^{\prime} \nonumber \\
 & = & -\sqrt{\sqrt{\pi}}\sum_n c_{0,n} \nu
\left[\int_{-\infty}^{\chi} P_n \sech^2 \chi^{\prime} \, \dif
\chi^{\prime} - \right. \nonumber \\
 & & \left. \int^{\infty}_{\chi} P_n \sech^2 \chi^{\prime} \, \dif
\chi^{\prime} \right],
\end{eqnarray}
where the $\tanh \chi$ argument of the Legendre polynomials has been
omitted for simplicity.

The integrals in the perturbed acceleration can be performed if we
take advantage of the following substitutions; $u=\tanh
\chi$, $\sech^2 \chi = 1-u^2$, $\dif u = (1-u^2)\, \dif \chi$.
The integrals involving the Legendre function become,
\begin{displaymath}
\int P_n(s) \, \dif s = \frac{P_{n+1}(s) - P_{n-1}(s)}{2n+1},
\end{displaymath}
which reduces Equation~\ref{pcbe} to
\begin{eqnarray}\label{pcbe4}
\lefteqn{\frac{\partial \fst_1}{\partial \tau} + \varpi (1-u^2) \frac{\partial
\fst_1}{\partial u} - u \frac{\partial \fst_1}{\partial
\varpi} -} \nonumber \\ & & \varpi \sqrt{\sqrt{\pi}} \left\{
\sum_n c_{0,n} \frac{1}{\nu} \left[P_{n+1}(u) - P_{n-1}(u)\right]
\right\} \times \nonumber \\ 
 & & (1-u^2) e^{-\varpi^2}=0.
\end{eqnarray}

Expanding the remaining $\fst_1$ functions in Equation~\ref{pcbe4}
with Equation~\ref{hlex} and using the orthogonality of the Hermite
and Legendre functions produces the following recursion relation
version of the collisionless Boltzmann equation,
\begin{eqnarray}\label{cbelp}
\dot{c}_{m,n} & = & L_{m,n}^{m-1,n-1} \, c_{m-1,n-1}
  + L_{m,n}^{m-1,n+1} \, c_{m-1,n+1}\nonumber\\ & + & L_{m,n}^{m+1,n-1}
  \, c_{m+1,n-1}  + L_{m,n}^{m+1,n+1} \, c_{m+1,n+1}.
\end{eqnarray}
The factors $L_{m,n}^{i,j}$, which can be arranged as matrix elements,
are given by 
\begin{eqnarray}\label{Ldef}
L_{m,n}^{m-1,n-1}&=&\frac{\sqrt{m}(n-1)n-2\delta_{1,m}}
{\sqrt{2(2n+1)(2n-1)}},\nonumber \\
L_{m,n}^{m-1,n+1}&=&-\frac{\sqrt{m}(n+2)(n+1)-2\delta_{1,m}}
{\sqrt{2(2n+1)(2n+3)}},\nonumber \\
L_{m,n}^{m+1,n-1}&=&\frac{\sqrt{m+1}(n+1)n}
{\sqrt{2(2n+1)(2n-1)}},\nonumber \\ 
L_{m,n}^{m+1,n+1}&=&- \frac{\sqrt{m+1}(n+1)n}{\sqrt{2(2n+1)(2n+3)}},
\end{eqnarray}
where $m,n,i,j \ge 0$.  In a situation that does not involve the
gravitational pull of the perturbation on equilibrium, one can imagine
test particles moving in a perturbed potential.  For such a case, the
recursion relation is obtained by omitting the Kronecker
$\delta_{1,m}$ terms.

\subsection{Non-linear Perturbations}\label{apert}

For large amplitude perturbations, the approach is similar to that
for linear perturbations.  We again decompose a distribution function
into Hermite and Legendre polynomials,
\begin{equation}\label{ahlex}
\fst=\sum_{m,n}  A_{m,n} \mu \nu
H_m(\varpi) P_n(\tanh \chi) \sech^2 \chi e^{-\varpi^2}.
\end{equation}
The significant difference from the linear case is the $\alpha
\partial \fst/\partial \varpi$ term in Equation~\ref{cbe2}.  In the
non-linear case, a perturbation will act on itself as well.
The result of this self-interaction is that a non-trivial triple
product of Legendre polynomials appears.  In the linear case, one of
the Legendre functions in the triple product is $P_1(u)=u$ and the
product can be handled more simply.  We take advantage of the fact
that,
\begin{displaymath}
P_j P_k = \sum_{s=0 \atop {\rm even}}^{j+k} Q_s^{(j,k)} P_{j+k-s},
\end{displaymath}
to reduce any triple product to a product that can be simplified using
the Legendre orthonormality relationship.  The $Q$ functions are
defined by \citep{d53},
\begin{displaymath}
Q_s^{(j,k)} = \frac{2j+2k-2s+1}{2j+2k-s+1} \frac{\lambda_{s/2}
\lambda_{j-s/2} \lambda_{k-s/2}}{\lambda_{j+k-s/2}},
\end{displaymath}
where
\begin{displaymath}
\lambda_B = \frac{(2B)!}{2^B (B!)^2},
\end{displaymath}
if $B \ge 0$ and is zero otherwise.

With this complication, the recursion relation version of
Equation~\ref{cbe2} expands to,
\begin{eqnarray}\label{cbeap}
\dot{A}_{m,n} & = & R_{m,n}^{m-1,n-1} \, A_{m-1,n-1}
  + R_{m,n}^{m-1,n+1} \, A_{m-1,n+1}\nonumber\\ & + & R_{m,n}^{m+1,n-1}
  \, A_{m+1,n-1}  + R_{m,n}^{m+1,n+1} \, A_{m+1,n+1} \nonumber\\
 & - & S_1 + S_2,
\end{eqnarray}
where
\begin{eqnarray}
\lefteqn{S_1= 2\sqrt{\sqrt{\pi}m(2n+1)} \times}
\nonumber \\
& & \sum_{i \ge 1}^{\infty}
\frac{A_{0,i}}{\sqrt{2i+1}}\sum_{s=0 \atop {\rm even}}^{n+i+1}
\frac{A_{m-1,n+i+1-s}}{2(n+i+1-s)+1} Q_s^{(n,i+1)},
\end{eqnarray}
and
\begin{eqnarray}
\lefteqn{S_2= 2\sqrt{\sqrt{\pi}m(2n+1)} \times}
\nonumber \\
& & \sum_{i \ge 1}^{\infty}
\frac{A_{0,i}}{\sqrt{2i+1}}\sum_{s=0 \atop {\rm even}}^{n+i-1}
\frac{A_{m-1,n+i-1-s}}{2(n+i-1-s)+1} Q_s^{(n,i-1)}.
\end{eqnarray}
The matrix elements $R_{m,n}^{i,j}$ are just the test-particle
versions of the $L_{m,n}^{i,j}$ for the linear case;
\begin{eqnarray}\label{Rdef}
R_{m,n}^{m-1,n-1}&=&\frac{\sqrt{m}(n-1)n}
{\sqrt{2(2n+1)(2n-1)}},\nonumber \\
R_{m,n}^{m-1,n+1}&=&-\frac{\sqrt{m}(n+2)(n+1)}
{\sqrt{2(2n+1)(2n+3)}},\nonumber \\
R_{m,n}^{m+1,n-1}&=&\frac{\sqrt{m+1}(n+1)n}
{\sqrt{2(2n+1)(2n-1)}},\nonumber \\ 
R_{m,n}^{m+1,n+1}&=&- \frac{\sqrt{m+1}(n+1)n}{\sqrt{2(2n+1)(2n+3)}}.
\end{eqnarray}
The Kronecker delta terms of the linear case are simply single terms
from the $S_1$ and $S_2$ sums when $s$ has its maximum value.

\section{Linear Perturbation Time-independent Solutions}

Based on the results of Section~\ref{lpert}, we next discuss routes to
steady states of the linearized collisionless Boltzmann equation in
test-particle and self-gravitating regimes using coefficient recursion
relations.  For test particles, phase-mixing will be the only
process active in the phase-space evolution of the system.  As a
result, any evolution will occur on the time-scale of phase-mixing.
In the self-gravitating case, violent relaxation will occur as
well.  The last term on the left-hand side of Equation~\ref{pcbe}
makes the system self-gravitating, and the perturbed acceleration is
the only term that can relate to this kind of relaxation.

The collisionless Boltzmann equation possesses an infinite set of
steady-state solutions, as does its linearized version \citep{bt87}.
In the following, we describe a procedure to obtain the general
solution of the steady-state linear problem for perturbations of the
separable equilibrium. The analysis yields a set of time-independent
modes which form a complete orthonormal basis that span a sub-space of
all possible configurations.  These modes can then be used to
construct any steady state.  More importantly, the projection of an
arbitrary small-amplitude initial perturbation onto this basis
produces the steady state that would result from evolving the system
according to Equation~\ref{pcbe}.

\subsection{Recursion Relation Procedure}\label{lprrp}

The form of the time-independent modes of the linear problem is
suggested by the solutions of the test-particle case, which have the
following form,
\begin{displaymath}
\fst(\beta \epsilon) = \fst(\varpi^2 + 2\phi_0(\chi)),
\end{displaymath}
where $\phi_0=\log(2\cosh\chi)$ is the external potential. For small
deviations from equilibrium, the distribution function can be expanded
in a power series in $\beta \epsilon$ times a Boltzmann kernel,
\begin{eqnarray} 
\lefteqn{\fst(\beta \epsilon) = \fst_0(\chi,\varpi)+\sum_{k=0}^\infty a_k (\beta
\epsilon(\chi,\varpi))^k e^{-\beta \epsilon}}  \nonumber \\ 
&=&\left[ \frac{1}{2\sqrt{\pi}}+\sum_{k=0}^\infty a_k \left(
\varpi^2 + 2\phi(\chi)\right)^k \right] e^{-\varpi^2} \sech ^2\chi,
\end{eqnarray}
where the $a_k$ are time-independent coefficients.

The set of linearly independent functions $\{\beta
\epsilon(\chi,\varpi)^k\}$ can be rendered into an orthonormal basis
$\{F^{(k)}(\chi,\varpi)\}$ via a Gram-Schmidt process, where the
test-particle $\{F^{(k)}(\chi,\varpi)\}$ are $k$th order polynomials in
$\varpi^2$ and $\phi_0(\chi)$.  Likewise, the solutions of the
self-gravitating Boltzmann equation can be written as
\begin{equation}\label{bkdf}
\fst(\chi,\varpi) = \fst_0(\chi,\varpi)+\sum_{\substack{k=2 \\k \;
{\rm even}}}^\infty b_k F^{(k)}(\chi,\varpi) e^{-\varpi^2}\sech^2\chi.
\end{equation}
The $b_k$ coefficients define the relative strengths of the various
time-independent mode contributions to the perturbation distribution
function.

We can further break the orthonormal $F^{(k)}$ functions into combinations
of velocity and position polynomials,
\begin{equation}\label{deffk}
F^{(k)}(\chi,\varpi) =\sum_{m=0}^k \frac{1}{\sqrt{2^m m!}} 
H_m(\varpi)G_{k,m}(\chi), 
\end{equation}
where the $G_{k,m}(\chi)$ can be written in terms of Legendre
polynomials,
\begin{equation}\label{defgkm}
G_{k,m}(\chi) =\sum_{n=0}^\infty \sqrt{2n+1} d^{(k)}_{m,n} 
P_{n}(\tanh \chi).
\end{equation}
For a given value of $k$, the $d^{(k)}_{m,n}$ coefficients obey the
recursion relation in Equation~\ref{cbelp}.  The difference between
test-particle and self-gravitating coefficient recursion relations is
what distinguishes the $F^{(k)}$ functions for the two cases.  Since we
are now looking at time-independent solutions, the time derivative
term in Equation~\ref{cbelp} must be set to zero.
As an example, a test-particle simulation will have $d$-coefficient
values that must obey this recursion relation,
\begin{eqnarray}\label{ctimeind1}
d^{(k)}_{m-1,n+1} & = & \sqrt{\frac{2n+3}{2n-1}} 
\frac{n(n-1)}{(n+2)(n+1)} d^{(k)}_{m-1,n-1}+ \nonumber \\
 & & \sqrt{\frac{(m+1)(2n+3)}{m(2n-1)}} \frac{n}{n+2} 
     d^{(k)}_{m+1,n-1} - \nonumber \\
 & & \sqrt{\frac{m+1}{m}} \frac{n}{n+2} d^{(k)}_{m+1,n+1}.
\end{eqnarray}

\begin{figure}
\includegraphics{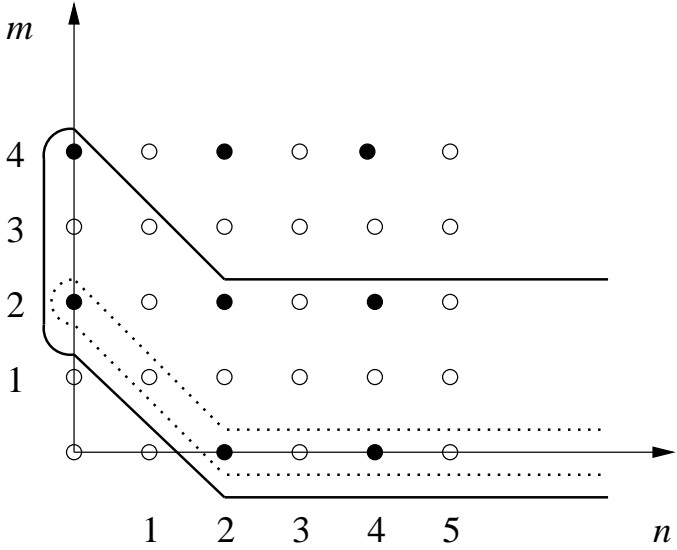}
\caption{Lower corner of the $m,n$ plane illustrating the groupings of
coefficients that make up time-independent modes.  The solid dots
indicate coefficients that can be non-zero.  Coefficients with odd
parity (either $m$ or $n$ is odd) are unpopulated as they give rise to
systems with non-zero centers-of-mass positions and velocities.
Coefficients with both odd $(m,n)$ values cannot be part of
time-independent solutions as they would phase mix until the only
non-zero coefficients exist only at very large $(m,n)$ values
($m=n=\infty$ in the $t=\infty$ limit).  The dotted line boundary
shows the coefficients involved in the first time-independent mode
$F^{(2)}$, while the solid boundary indicates those coefficients
linked in the second time-independent mode $F^{(4)}$.
\label{mngrid}}
\end{figure}

The problem of solving the time-independent Boltzmann equation is
transformed into solving recursion relations on the $(m,n)$ grid in
the region $0\le m\le k$, $n\ge 0$, where the $m=k$ row is set to
zero, except $d^{(k)}_{k,0}$ which is left as a free parameter.  Any
coefficient with an odd $m$ or $n$ index must be zero as those
coefficients give rise to non-zero center-of-mass position and/or
velocity values that are incompatible with a time-independent state.
Figure~\ref{mngrid} illustrates an example layout of a
$d^{(k)}$-coefficient grid.  The coefficients on the left-most column
$d^{(k)}_{m,0}$ are likewise left as free parameters for the
subsequent Gram-Schmidt orthogonalization procedure.  For a given $k$,
one starts at the upper left-hand corner ($m=k$) and uses the
appropriate recursion relation to find the coefficients for
$(m=k-2,n>0)$ up to some $n_{max}$, working left to right.  The
cut-off $n_{\rm max}$ is chosen so that the series for $G_{k,m}$ is
well-behaved (see \S~\ref{limit}).  Once the $m=k-2$ row has been
completed, the process can be repeated for all $k-2 > m >0$, working
downward.  Once the coefficients for each $k$ are determined, the free
parameters are used to construct an orthonormal basis via a
Gram-Schmidt process.

This is essentially the path we follow, with a few important details
to be added.  Thinking of the dynamics problem in general, we re-cast
the coefficient Boltzmann equation (Equation~\ref{cbelp}) as a matrix
equation.  First, arrange the $(0\le m \le k,0 \le n \le n_{\rm max})$
$\dot{d}^{(k)}_{m,n}$ and $d^{(k)}_{m,n}$ terms involved in
Equation~\ref{cbelp} as vectors.  The $L^{i,j}_{m,n}$ factors can then
be organized into a two-dimensional matrix.  This results in the
following relationship,
\begin{equation}\label{matcbe}
\mathbfss{L}\bmath{d}^{(k)}=\bmath{\dot{d}}\ \hspace{-0.2em}^{(k)}.
\end{equation}
Assuming that the time-dependence of a $\bmath{d}^{(k)}$ is given
by $\exp{(\lambda^{(k)} t)}$, then Equation~\ref{matcbe} transforms
into an eigenvalue equation,
\begin{equation}\label{matcbe2}
\mathbfss{L}\bmath{d}^{(k)}= \lambda^{(k)} \bmath{d}^{(k)}.
\end{equation}
Solving this equation is straightforward, but a complication arises
that impacts any subsequent Gram-Schmidt process.  The matrix
$\mathbfss{L}$ is not symmetric.  This means that there are so-called
left- and right-handed eigenfunctions (sets of $d^{(k)}_{m,n}$), but
left/right eigenfunctions are only orthogonal to their opposite-handed
counterparts.  As the Gram-Schmidt process relies on orthogonal
functions, we need to find both right- and left-handed eigenfunctions
for $\lambda^{(k)}=0$.  This means we must also find eigenfunctions of
$\mathbfss{L}^{\rm T}$.  With both left- and right-handed
eigenfunctions, the Gram-Schmidt process can proceed as long as
left/right pairs of functions are used for orthonormality.  We note
that the same approach could be taken to determine time-dependent
solutions ($\lambda^{(k)}\ne 0$).  A full discussion of such solutions
will be postponed to maintain focus on the process of determining
steady states, however the general behavior of any $(m,n)$
perturbation is to couple to higher-index coefficients \citep{br14}.
Essentially, any initial condition contains an infinite number of
time-dependent modes.  These modes then quickly phase mix, leaving
only the time-independent modes as the visible remnant of the initial
conditions.

As examples of the procedure, we consider the determination of the
first two time-independent modes.  The first mode is built from an
initially unknown normalization constant $d^{(2)}_{2,0}$.  The right-
and left-handed coefficient sets are found by applying the
$\mathbfss{L}$ and $\mathbfss{L}^{\rm T}$ recursion relation
operations, respectively.  Recall that in our scheme, all coefficients
with $(m \ge 2, n > 0)$ are zero at this stage -- only the
coefficients with $m=0$ and $n \ge 2$ ($n$ even) will be non-zero.
Equation~\ref{ctimeind1} with $m=1,n=1$ gives the link between the
normalization constant and the first $m=0$ coefficient.  For $n>1$,
the recursion relation of Equation~\ref{ctimeind1},
\begin{equation}
d^{(2)}_{0,n+1}=\sqrt{\frac{2n+3}{2n-1}} \frac{n(n-1)}{(n+2)(n+1)}
d^{(2)}_{0,n-1},
\end{equation}
links the remaining coefficient values, which are then known in terms
of the normalization constant.  The value of this constant is
determined by the orthogonality relationship between the left- and
right-handed coefficients,
\begin{eqnarray}
\lefteqn{\int \int \sum_{m,n} [d^{(k,L)}_{m,n} H_m(\varpi) P_n(\tanh
\chi)] \times} \nonumber \\
& & [d^{(k,R)}_{m,n} H_m(\varpi) P_n(\tanh \chi)] e^{-\varpi^2}
\sech{\chi}^2 \dif \varpi \dif \chi = 1,
\end{eqnarray}
where the $L$ and $R$ superscripts on the coefficients indicate their
handedness.  At this point, we have a time-independent mode (set of
coefficients) that is orthonormal to equilibrium.

To continue, we allow for two initially undetermined constants,
$d^{(4)}_{4,0}$ and $d^{(4)}_{2,0}$.  We again apply the
$\mathbfss{L}$ and $\mathbfss{L}^{\rm T}$ recursion relations to
determine right- and left-handed coefficient sets in terms of the
uknown constants.  At this stage all coefficients with $(m \ge 4, n >
0)$ are zero.  The orthogonality relation between the first and second
mode coefficient sets,
\begin{eqnarray}
\lefteqn{\int \int \sum_{i,j} \sum_{m,n} [d^{(k^{\prime},L)}_{i,j}
H_i(\varpi) P_j(\tanh \chi)] \times} \nonumber \\ 
& & [d^{(k,R)}_{m,n} H_m(\varpi) P_n(\tanh \chi)] e^{-\varpi^2} 
\sech{\chi}^2 \dif \varpi \dif \chi = 0,
\end{eqnarray}
allows us to determine the $d^{(4)}_{2,0}$ value.  Finally, the
normalization condition provides us with the condition to find
$d^{(4)}_{4,0}$.  This procedure continues similarly for higher-order
time-independent mode calculations.  Normalization provides one of the
free parameters, while orthogonality with the previous functions sets
the remainder.

The importance of these time-independent modes lies in the ability to
predict a steady state based on initial conditions.
If a system is gently perturbed from the $f_0$ equilibrium, the
perturbations evolve according to the linearized Boltzmann equation by
dephasing in phase-space and, if self-gravity is present, by
transferring particles into and out of equilibrium.  At any point in
time in an evolution, one can imagine the system being composed of a
steady-state component and a decaying, fluctuating component.  In
the linear problem, the final state can be predicted by projecting the
initial conditions into the time-invariant sub-space.
A time-independent coefficient $b_k$ is
found simply by taking the inner product of the initial conditions
with $F^{(k)}$,
\begin{equation}\label{genbk}
b_k=\int f(\chi,\varpi,t=0) F^{(k)}(\chi,\varpi) \dif \chi \dif \varpi.
\end{equation}
For $N$-body distribution functions composed of delta functions, the
coefficients can be calculated from the average value of $F^{(k)}$,
\begin{equation}\label{nbbk}
b_k=\left< F^{(k)} \right> =\frac{1}{N}\sum_{i=1}^{N}
F^{(k)}(\chi_i,\varpi_i).
\end{equation}

Note that the $F^{(k)}$ functions involved in Equations~\ref{genbk} and
\ref{nbbk} can be either left- or right-handed, as they depend on the
likewise handed $d^{(k)}_{m,n}$ values through Equations~\ref{deffk} and
\ref{defgkm}.  The handedness chosen for the $F^{(k)}$ in these
expressions must be opposite to that assumed for the distribution
function in Equation~\ref{bkdf}.  For concreteness in what follows, we
have assumed that the perturbation distribution function is composed
of right-handed $F^{(k)}$ functions, which then demands that the
orthonormal left-handed $F^{(k)}$ be used to calculate time-independent
mode coefficients.

\subsection{Limitations}\label{limit}

\begin{figure}
\scalebox{0.5}{
\includegraphics{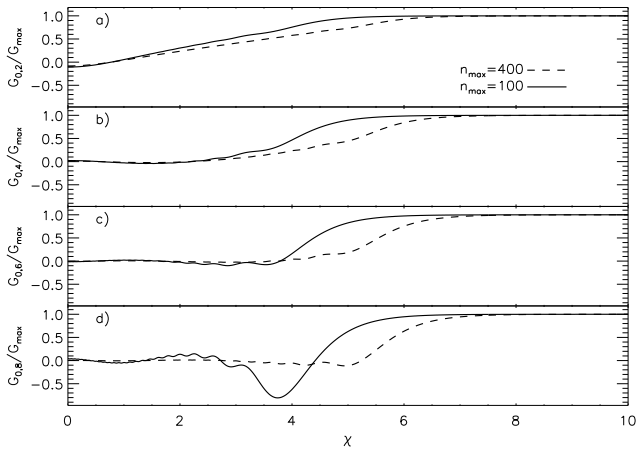}}
\caption{Approximations to $G_{k,0}$ curves for $k=2,4,6,8$.  The
exact $G_{k,m}$ function involves an infinite series of Legendre
functions, while the approximations shown here are from a series
truncated at a Legendre indexes $n_{\rm max}=100$ and $n_{\rm
max}=400$. These curves are closely related to the densities of the
various time-independent modes.  Note that the high frequency
oscillations become more apparent as $k$ increases.  These
oscillations decrease in magnitude as the value of $n_{\rm max}$ is
increased.
\label{converge}}
\end{figure}

In order for our Gram-Schmidt approach to be implemented, we have to
truncate infinite series, \ie\ recursion relations need to be solved
over finite regions of coefficient space.  Another example where
truncation plays a role, the $G_{k,m}(\chi)$ functions of
Equation~\ref{defgkm} must be approximated using a finite number of
terms.  If we focus on only $m=0$ terms, we are looking at the
essential behavior of densities associated with the various
time-independent modes.  The actual densities of the modes involve
multiplying by a $\sech^2 \chi$ term, so the behavior at large $\chi$
is effectively killed.  Figure~\ref{converge} shows the impact of
changing mode and maximum $n$ value on these curves.  Through trial
and error, we have settled on $n_{\rm max}=400$ as an acceptable limit
for this work.  We have also set $k_{\rm max}=16$ as the range of
time-independent modes created.  In the end, these limits have been
adopted because of the success the scheme has had in describing the
results of $N$-body simulations (see Section~\ref{lsims}).  

\subsection{Analytical Comparison}

As a check on the scheme described above, we compare our set of
$d^{(2)}_{m,n}$ with coefficients found through a different
route.  We imagine changing the temperature of an equilibrium, related
to $\beta$ from Section~\ref{intro}, by a small amount.  This should
produce another equilibrium, which should be time-independent.
Thinking of this as a series expansion,
\begin{equation}
\fst_0(\beta + \delta) = \fst_0(\beta) + \delta \frac{\partial
\fst_0}{\partial \beta},
\end{equation}
to first order in the temperature change.  The derivative is
\begin{equation}
\frac{\partial \fst_0}{\partial \beta}=-\frac{\fst_0}{\beta}
\left[\left( \varpi^2- \frac{3}{2} \right) +2\chi \tanh{\chi} \right].
\end{equation}
The term in square brackets is the perturbation distribution function.
It is straightforward to show that this distribution function is
time-independent, as intended.  We have calculated the $(m=0,n)$
coefficient values that correspond to this perturbation.  They are not
normalized in the same way as the $d^{(2)}_{m,n}$ values, but their
successive values have the same ratios ($m=0,n=2$ over $m=0,n=4$, for
example) as for the $d^{(2)}_{0,n}$.  In other words, our recursion
relation approach reproduces a known time-independent mode.

\subsection{Energy Characteristics}

With the linear time-independent modes identified, we next report on
their energy properties.  Here, we focus on self-gravitating
situations, as test particle systems do not have interesting energy
behaviors.  The kinetic energy content of any mode is determined
solely by the value of the $d^{(k)}_{2,0}$ coefficient.  This simple
form results from the fact that calculating the second velocity moment
of any distribution function that is expanded as in
Equation~\ref{hlex} is non-zero only when $m=2$ and $n=0$.  The
potential energy content of any mode involves only coefficients with
$m=0$ and even $n>0$, the same as acceleration \citep{br14}.

With the $d^{(k)}_{m,n}$ coefficients from above, we find that only
the $F^{(2)}$ mode contains energy.  This is reasonable, as this mode
corresponds to changing the temperature of the system.  All other
$F^{(k)}$ with $k \ge 4$ have kinetic and potential energies that are
equal in magnitude and opposite in sign.  The values also indicate
that all $F^{(k)}$ modes are in virial equilibrium.  Again, this is
unsurprising as it must be time-independent.  The connection with
these functions and energy lead us to refer to these functions as $E$
modes.

For general linear perturbations, there will be some energy-bearing
component and some non-energetic components.  It is important to note
that the non-energetic components can still affect the spatial and velocity
density of a system.  As a result, the structure of a steady state
composed of a combination of time-independent modes is uniquely
determined by its initial conditions, and not by some general
principle such as entropy maximization.

\subsection{Alternative Approaches}

The scheme we have laid out here is not unique.  It is convenient
because the coefficients involved in each mode have simple links to
quantities like kinetic and potential energy.  However, one could
choose to form different time-independent modes.  For example, if all
coefficients with $m>2,n>2$ are set equal to zero and $d_{0,2}$ is
left as an undetermined constant, then the recursion relations can be
used to calculate $d_{m,0}$ values, for $m \le m_{\rm max}$.  In
essence, Figure~\ref{mngrid} could be flipped about the $m=n$
diagonal.

The recursion relations change for this approach, but the analogue to
$\mathbfss{L}$ remains non-symmetric.  As a result, both left- and
right-handed eigenfunctions must be determined as before, and the
Gram-Schmidt technique involving both functions must be employed.  To
contrast with the $E$ modes described above, we refer to these
alternative functions as $B$ modes.  As distinction from the $b_k$
values for $E$ modes, the time-independent coefficients related to $B$
modes are labeled as $y_k$.  

\subsection{Simulations}\label{lsims}

Test particle evolutions in the equilibrium potential use an adaptive
time step, Runge-Kutta scheme to track particles.  Particle
accelerations are determined by the equilibrium potential only.
Tolerances are chosen so that the total energies of test particle
systems experience fractional variations on the order of $10^{-11}$.

When allowing the perturbation to self-consistently evolve,
substantially more care must be taken with a simulation.  Fortunately,
the distance-independence of the gravitational force in the
one-dimensional problem allows one to take advantage of a key
simplification.  During an evolution, all particles move with constant
acceleration between crossings.  As a result, kinematic equations
precisely predict the motions of particles (only numerical round-off
errors degrade the process) and total system energies vary by
approximately $10^{-9}$ during thousand-particle evolutions over a
thousand dynamical times.

Rather than setting a fixed time step for numerical evolution, the
conditions of the simulation determine when particle positions and
velocities are updated.  Initially, the time until the next collision
of nearest neighbor pairs is calculated for every pair.  The pair with
the shortest interval sets the time step and the two particles that
exist at a common location switch their constant acceleration values
as they pass one another.  By keeping track of a particle's last and
next crossing times, only a few particle must be updated after a
time step.  The bookkeeping is made easier when crossing times are
stored in a heap structure that can be quickly re-sorted
\citep{n03,jw11}.

Initial conditions for simulations are created by exciting specific
modes to perturb the separable equilibrium.  In practice, this process
requires some caution during implementation.  Perturbing modes can
involve negative distribution function values, at least for some
values of $\chi$ and $\varpi$.  Since we do not have a simple way of
incorporating negative masses into our simulations, care must be taken
with the amplitudes of any such modes.  If one cavalierly assigns a
single mode amplitude, other, unintended modes can appear in the
following manner.  Any generated $N$-body initial distribution
function is max[0,$\fst(\chi,\varpi)$], not $\fst(\chi,\varpi)$, which
can be negative.  The unintended modes are those that are needed to
make $\fst \ge 0$.  Unless otherwise noted, we have fixed
perturbations strengths at values that render this problem negligible.

Our simulations consist of ensembles of 100 distinct realizations of a
given initial distribution function, each with $N=1024$ particles.  Each
realization is evolved independently and ensemble-averaged quantities
are then created.  Typically, evolutions end at $\tau=5T$, where $T$
is the crossing time-scale for a constant-density system with mass
$\mt$.  That this time range is adequate to guarantee that simulations
reach steady states will be made clear in the following discussion.
Our evolution code tracks quantities like energies (kinetic,
potential, total), coefficient values, and entropy.  Entropy in these
$N$-body simulations is calculated using a particle counting scheme,
\begin{equation}
S = - \sum_i n_i \ln{n_i},
\end{equation}
where $n$ is the particle count and $i$ enumerates different areas of
phase space (all of size $\Delta \chi \Delta \varpi$).  Unlike in
quantum situations where $\Delta \chi$ and $\Delta \varpi$ can be
related to Planck's constant, we have simply used trial and error to
set sizes of the phase-space boxes.  After investigating a wide range,
we have found that values near the adopted $\Delta \chi = \Delta
\varpi = 2\times 10^{-2}$ produce entropy values that show the most
obvious changes during evolution.  Smaller values result in almost no
particles falling into the boxes, while larger values produce boxes so
large that variation is basically absent.  In either case, resulting
entropy changes are small.

To investigate how well our time-independent modes describe steady
states, we have run several suites of simulations.  Initial conditions
consist of simple perturbations to equilibrium due to single
coefficients; $c_{2,0}$, $c_{0,2}$, and $c_{2,2}$.  Note that these
coefficients are distinct from the time-independent coefficients we
denote by $d^{(k)}_{m,n}$.  For each perturbation, we have varied the
strength, $0.05 \le \Delta \le 0.30$.  In this way, we map the range
of steady states that are well-described by the time-independent
modes.

\begin{figure}
\scalebox{0.5}{
\includegraphics{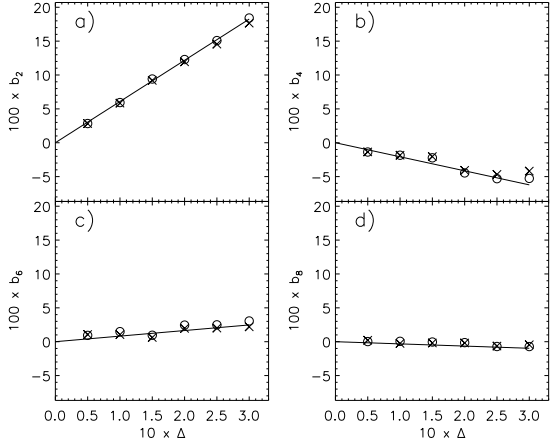}}
\caption{
The behavior of time-independent $E$ mode coefficients as the strength
of a self-gravitating $c_{0,2}$ perturbation is varied.  Panels a, b,
c, and d contain the first four coefficients, respectively.  In each
panel, the solid lines show the predicted behavior based on our
Gram-Schmidt orthogonalization scheme.  The results of simulations are
shown by the symbols.  Open circles represent the initial ensemble
average values of the coefficients, while the crosses show the final
values.
\label{c02tis}}
\end{figure}

\begin{figure}
\scalebox{0.5}{
\includegraphics{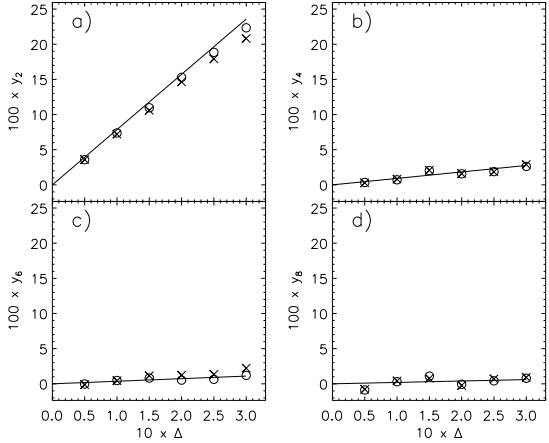}}
\caption{
The behavior of time-independent $B$ mode coefficients as the strength of
a self-gravitating $c_{0,2}$ perturbation is varied.  Panels and
symbols are analogous to those in Figure~\ref{c02tis}.  While these
alternative $B$ modes provide an acceptable basis for analyzing steady
states, their lack of connection to physical quantities, like
energy, make them less appealing than the $E$ modes.
\label{c02btis}}
\end{figure}

Figure~\ref{c02tis} compares predictions from our time-independent $E$
mode coefficient sets to the outcome of self-gravitating simulations
with $c_{0,2}$ perturbations.  To simplify this discussion, unless
otherwise specified, time-independent coefficients discussed will be
those for $E$ modes.  Solid lines show how the $b_k$ coefficients
should vary with $\Delta$ for linear perturbations.  These examples
extend only to $k=8$ for brevity, but similar plots up to $k=16$ show
similar levels of agreement.  Note that we have multiplied the
perturbation strengths by 10 for the horizontal axes and the
coefficient values by 100 for the vertical axes.  The open circles are
centered at the initial values of the coefficients while the crosses
indicate final values.  In general, there is good agreement between
the predictions and simulated results.  Unsurprisingly, as the
perturbation strength grows, the separation between the initial and
final values grows.  Figure~\ref{c02btis} is the $B$ mode analogue to
Figure~\ref{c02tis}.  As with the $E$ modes, the $B$ mode coefficients
derived from simulations match predictions well.

\begin{figure}
\scalebox{0.5}{
\includegraphics{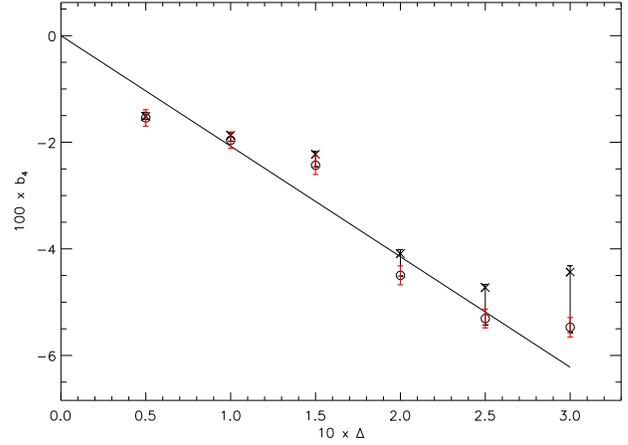}}
\caption{A more focused view of the same information in panel b of
Figure~\ref{c02tis}.  The line and symbols have the same meanings as
in Figure~\ref{c02tis}.  The thick error bars show the size of the
error-in-the-mean for an ensemble.  The thin error bars show the full
range of coefficient values for an ensemble.
\label{c02tisz}}
\end{figure}

\begin{figure}
\scalebox{0.5}{
\includegraphics{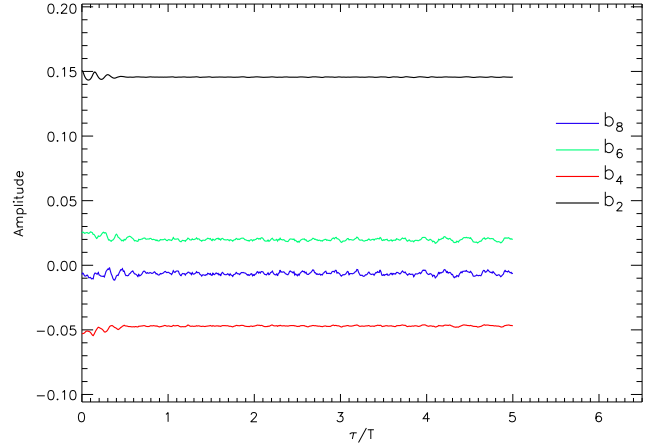}}
\caption{Time evolution of time-independent mode coefficients
determined from an ensemble of self-gravitating simulations with an
initial $c_{0,2}$ perturbation with strength $\Delta=0.25$.  The
variations in coefficient values visible here indicate that there is
some amount of non-linearity present in these simulations.
\label{c02bkt}}
\end{figure}

Figure~\ref{c02tisz} is a focused version of panel b from
Figure~\ref{c02tis}.  In this figure, the thick error bars show the
error-in-the-mean range.  The thin error bars represent the full
range of coefficient values for an ensemble.  The changes in
coefficient values during an evolution and the overall ranges in the
coefficient values in an ensemble both grow with perturbation
strength.  The linear assumption underlying our prediction line is
breaking down at the highest perturbation strengths investigated here.
Figure~\ref{c02bkt} shows coefficient behaviors as a function of
time for a perturbation strength $\Delta=0.25$.  The changes in
coefficient values early in the evolution highlight the non-linearity
of this situation.

\begin{figure}
\scalebox{0.5}{
\includegraphics{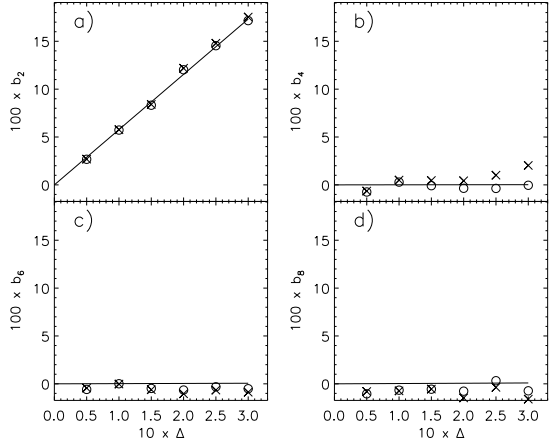}}
\caption{The time-independent mode coefficient behaviors from
simulations with initial $c_{2,0}$ perturbations.  Lines and symbols
represent the same quantities as in Figure~\ref{c02tis}.
\label{c20tis}}
\end{figure}

For comparison, the results of self-gravitating simulations with
initial $c_{2,0}$ perturbations are shown in Figure~\ref{c20tis}.  The
same basic agreement between predictions and simulations is evident,
and the discrepancies set in around the same perturbation strength as
previously noted.

\begin{figure}
\scalebox{0.5}{
\includegraphics{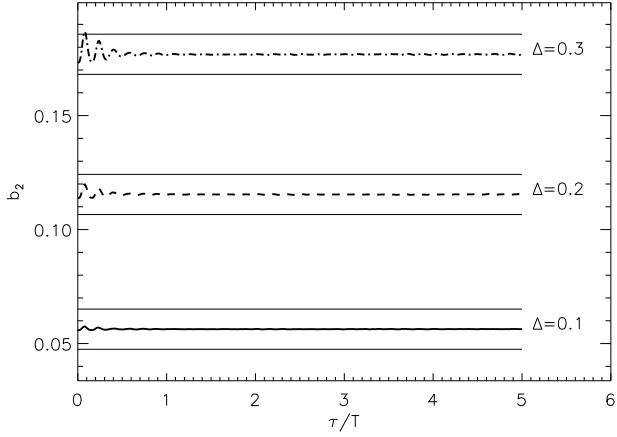}}
\caption{Thick lines show the time evolutions of time-independent mode
coefficients determined by ensemble averaging simulations with initial
$c_{2,0}$ perturbations.  Different line styles reflect the
perturbation strengths indicated.  The thin bounding lines show the
size of the ensemble error-in-the-mean range.  Unsurprisingly, $N$-body
non-linearities grow with perturbation strength.  However, at least
for these low order perturbations, the effects of non-linearities are
at worst comparable to statistical noise.
\label{c20nonlin}}
\end{figure}

Looking in more detail at the $b_2$ values resulting from a $c_{2,0}$
perturbation with different strengths shows how simulation
non-linearities impact the coefficient values.  Figure~\ref{c20nonlin}
shows time evolutions of ensemble-averaged $b_2$ coefficients over the
range of perturbations shown in Figure~\ref{c20tis}.  The thin lines
bounding the various evolutions indicate the ensemble
error-in-the-mean ranges for each set of simulations.  As the
perturbation strength increases, the non-linearity of the simulations
increases, but stays roughly within the statistical uncertainty of the
coefficients.  Higher time-independent-mode coefficient evolutions can
be noisier than those for $b_2$, but overall any non-linearity due to
the $N$-body nature of the simulations can be considered small.

\begin{figure}
\scalebox{0.5}{
\includegraphics{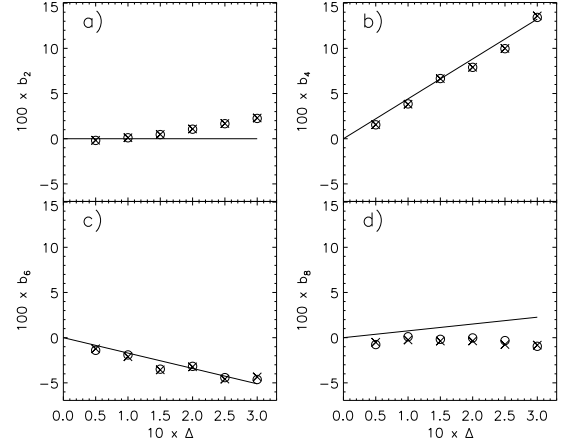}}
\caption{The time-independent mode coefficient behaviors from
simulations with initial $c_{2,2}$ perturbations.  Lines and symbols
represent the same quantities as in Figure~\ref{c02tis}.  The
appearance of first time-independent modes in panel a indicates that
the linear perturbation regime exists only for the lowest strengths
investigated here.
\label{c22tis}}
\end{figure}

As noted earlier, increasing perturbation strength can lead to
unintended modes being populated in a simulation.  A good example of
this occurs in self-gravitating simulations with initial $c_{2,2}$
perturbations.  In these systems, there should be no possibility of
having the first time-independent mode (with coefficient $b_2$).
Figure~\ref{c22tis} shows that for the lowest perturbation strengths,
this is reasonably achieved.  However, for even modest strengths
($\Delta =0.15$) we see this first time-independent mode appearing in
Figure~\ref{c22tis}a.  This is a consequence of higher $m,n$
perturbations causing negative distribution functions at lower
perturbation strengths.  Subsequently, our simulations leave the
linear regime for smaller $\Delta$ compared to those for $m=0,n=2$ and
$m=2,n=0$ cases.

\begin{figure}
\scalebox{0.5}{
\includegraphics{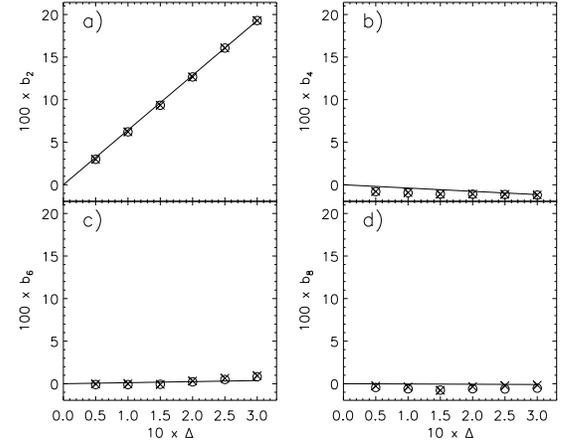}}
\caption{The time-independent mode coefficient behaviors from
test-particle simulations with initial $c_{2,0}$ perturbations.  Lines
and symbols represent the same quantities as in Figure~\ref{c02tis}.
This figure should be compared to its self-gravitating counterpart,
Figure~\ref{c20tis}.
\label{tc20tis}}
\end{figure}

\begin{figure}
\scalebox{0.5}{
\includegraphics{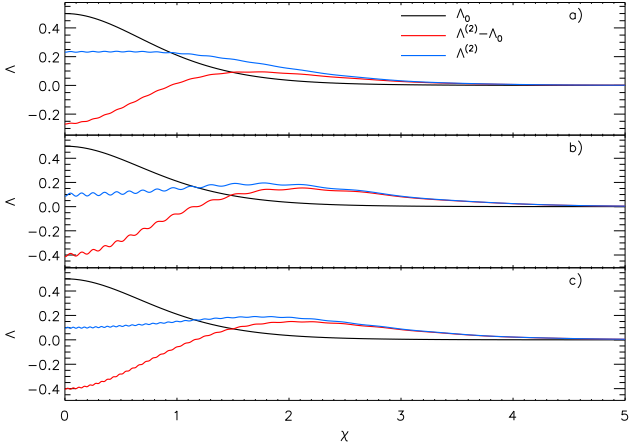}}
\caption{As in Figure~\ref{c20nonlin}, thick lines show the time
evolutions of time-independent mode coefficients determined by
ensemble averaging test-particle simulations with initial $c_{2,0}$
perturbations.  The thin lines still represent the ensemble
error-in-the mean ranges of the coefficient values.
\label{c20testnl}}
\end{figure}

All of the previous discussion has involved self-gravitating
simulations.  Analyses of test-particle simulations yield very similar
results.  Figures~\ref{tc20tis} and \ref{c20testnl} are the
test-particle analogues to Figures~\ref{c20tis} and \ref{c20nonlin},
respectively.  Note that even the small amount of non-linearity in the
self-gravitating case is absent.

\section{Non-Linear Time-independent Solutions}

As mentioned previously, we need to discuss families of
time-independent solutions for the non-linear perturbation case.
The specific modes identified for the linear case result because
time-independent coefficients values ($d^{(k)}_{2,0}$,
$d^{(k)}_{4,0}$, etc.) can be fixed via orthonormality.

Due to the non-linear nature of Equation~\ref{cbeap}, the procedure
for calculating time-independent $A$ coefficients changes.  First, the
highest $m$ row for the solution cannot be truncated after the $n=0$
term.  Second, an iterative approach needs to be taken.  This is
analogous to a relaxation approach to solve Poisson's equation in
two-dimensions \citep{press94}.  Families of solutions are
determined by choosing the maximum $m$ that will be allowed (denoted
by $k$ in analogy to the linear case), and family members are
distinguished by the value of $A_{k,0}$.  As with the linear
perturbation results, we will indicate the solution family with a
superscripted index, $A^{(k)}$.

As before, we use the specific example of the $A^{(2)}$ family to
illustrate the process.  Initially, $A_{2,0}^{(2)}$ is the only
non-zero coefficient.  All $A_{m>2,n}^{(2)}$ are zero and will remain
so.  Additionally, any odd-parity coefficients will be zero and fixed
as well.  We use the fact that $\dot{A}_{1,1}^{(2)}=0$ and
$\dot{A}_{3,1}^{(2)}=0$ to start finding time-independent
coefficients.  Equation~\ref{cbeap} with the conditions given results
in two equations that only involve $A_{0,2}^{(2)}$, $A_{2,0}^{(2)}$,
and $A_{2,2}^{(2)}$.  Again, any higher $n$ terms are assumed to be
zero at this point.  The two equations can be solved simultaneously,
giving first estimates of $A_{0,2}^{(2)}$ and $A_{2,2}^{(2)}$.  Next,
we use $\dot{A}_{1,3}^{(2)}=0$ and $\dot{A}_{3,3}^{(2)}=0$.  With our
$n=2$ coefficient estimates, we can solve the resulting equations for
$A_{0,4}^{(2)}$ and $A_{2,4}^{(2)}$.  In this manner, estimates for
the time-independent coefficients can be found up to some $n_{\rm
max}$.  Once the $n_{\rm max}$ passes are completed, the next
iteration begins again with $\dot{A}_{1,1}^{(2)}=0$ and
$\dot{A}_{3,1}^{(2)}=0$.  The now non-zero coefficients at higher $n$
values enter non-linearly and affect the new estimates of
$A_{0,2}^{(2)}$ and $A_{2,2}^{(2)}$.  Marching back out to $n_{\rm
max}$ likewise updates all other coefficients.  Repeating iterations,
the true time-independent coefficient values are approached.

In practice, we have also implemented the same kind of numerical
dissipation that one would use in a relaxation Poisson solver.  At the
end of an iteration, coefficient values are reset to the average of
the current and previous sets.  We have found that this technique
reduces required iterations by at least a factor of two, for a given
level of convergence.  After 10 iterations, we find that the average
change in coefficient values is less than one percent for the $m=k$
set and 1-2 orders of magnitude smaller for the lower $m$ sets.
Likewise, $\dot{A}_{m,n}^{(2)}=\mathcal{O}(10^{-6})$ for all $m$ and
$n$.  Finally, the virial ratio for our solutions, $2K/U$, is
typically $1+\varepsilon$, where $|\varepsilon|$ is $\mathcal{O}
(10^{-4})$, but this does increase with increasing $A_{2,0}^{(2)}$.

\subsection{Solution Behavior}

\begin{figure}
\scalebox{0.5}{
\includegraphics{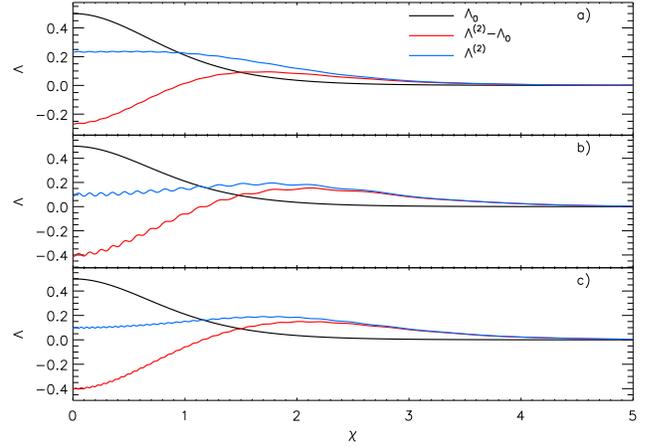}}
\caption{Plots of $A^{(2)}$ time-independent density distributions
compared to the equilibrium density, $\Lambda_0$.  In each panel, the
total density distribution is the line labeled $\Lambda^{(2)}$, while
the perturbation density is labeled $\Lambda^{(2)} - \Lambda_0$.
Perturbation strengths are given by $\Delta=A_{2,0}^{(2)}/A_{0,0}$.
In panel a, $\Delta=0.5$ with $n_{\rm max}=64$.  Panel b shows the
results of increasing to $\Delta=1.0$ while keeping $n_{\rm max}=64$.
The effect of increasing $n_{\rm max}$ is highlighted by comparing
panels b and c.  In panel c, $\Delta=1.0$ but $n_{\rm max}=128$.
\label{ati2dcomp}}
\end{figure}

Typically, $n_{\rm max}=64$, but values up to 256 have been used
successfully.  The major impact of increasing $n_{\rm max}$ is to
smooth the central regions of the distribution function.  The density
curves in Figure~\ref{ati2dcomp} highlight the impact of $n_{\rm max}$
and perturbation strength $\Delta = A_{2,0}^{(2)}/A_{0,0}$.  The top
panel compares the equilibrium density to a time-independent density
distribution with $\Delta=0.5$ and $n_{\rm max}=64$.  The small
variations seen near $\chi=0$ become magnified in the middle panel as
$\Delta$ increases to 1.0.  By increasing $n_{\rm max}$ to 128 in the
bottom panel, the central oscillations are made smaller.

\begin{figure}
\scalebox{0.5}{
\includegraphics{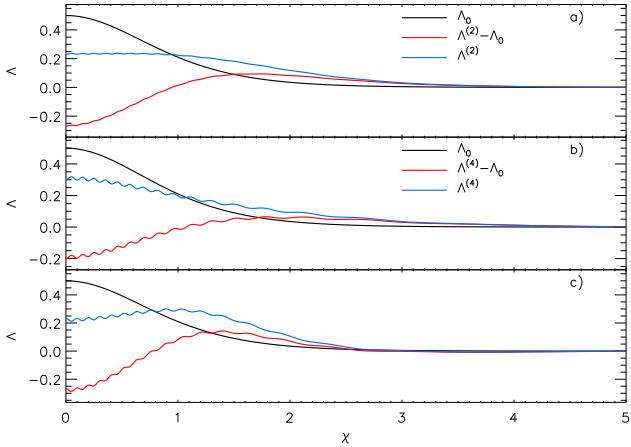}}
\caption{Plots of time-independent density distributions compared to
the equilibrium density, $\Lambda_0$.  As in Figure~\ref{ati2dcomp},
both the total and perturbation densities are shown.  All curves shown
result from $n_{\rm max}=64$.  Perturbation strengths are given by
$\Delta_2=A_{2,0}^{(4)}/A_{0,0}$ and $\Delta_4=A_{4,0}^{(4)}/A_{0,0}$.
For comparison, an $A^{(2)}$ with $\Delta=0.5$ is reproduced in panel
a.  Panel b shows an $A^{(4)}$ density with $\Delta_2=0.5$ and
$\Delta_4=0.25$.  The addition of the $m=4$ row has increased the
variations seen in the density.  Panel c shows a density distribution
of an $A^{(4)}$ member that has negative values for $\chi \ga 2.5$.
In this case, $\Delta_2=0.2$ and $\Delta_4=-0.4$.
\label{ati4dcomp}}
\end{figure}

We have extended this scheme to also create higher-order families.  As
an example, we describe how the scheme changes by examining the
$A^{(4)}$ family.  This is a two parameter family described by
$A_{2,0}^{(4)}$ and $A_{4,0}^{(4)}$.  With this family, three
simultaneous equations need to be solved; $\dot{A}_{1,1}^{(4)}=0$,
$\dot{A}_{3,1}^{(4)}=0$, and $\dot{A}_{5,1}^{(4)}=0$.  While we have
not investigated higher-order families in the same detail as the
$A^{(2)}$ and $A^{(4)}$, we have successfully found solutions for $k
\ge 6$ by solving $k/2 + 1$ equations simultaneously and following the
general procedure.  As with the $A^{(2)}$ family, we show a few
representative density profiles in Figure~\ref{ati4dcomp}.  Unlike the
linear perturbation case, the non-linear coupling in a large amplitude
perturbation makes it impossible to determine from initial conditions
which family the steady state will belong to.  However, the total
energy of the initial system could be used to select compatible family
members.  For example, an initial perturbation that makes the energy
differ from the equilibrium value must evolve to a steady state with
the same energy.  Calculating energies for $A^{(2)}$ and $A^{(4)}$
solutions with positive kinetic energies and non-negative density
distributions (see \S~\ref{bound}) reveals that there are one-to-one
correspondences between energy values and $\Delta$ values.  For the
$A^{(2)}$ family, the energy follows $\epsilon \approx 1.5 +
2.1\Delta$ from the equilibrium value $\epsilon=1.5$.  With the
$A^{(4)}$ family, this becomes $\epsilon \approx 1.5+ 2.1\Delta_2 -
0.01\Delta_4$.  This weak dependence on $\Delta_4$ reflects that while
the velocity distributions of these solutions vary substantially,
their potential energies are nearly the same.  For a fixed energy, one
can determine the relationship between $\Delta_2$ and $\Delta_4$ that
the steady state must have.

As previous work has found that the Lynden-Bell distribution function
has some success in describing steady states \citep{jw11}, we note
that none of the $A^{(2)}$ family members investigated here closely
resemble the Lynden-Bell form as $A^{(2)}$ densities have more
extended core structure.  As a quick summary of the Lynden-Bell
distribution function, it serves as a distinguishable particle
counterpart to the Fermi-Dirac distribution function \citep{lb67},
\begin{displaymath}
f_{\rm LB}(\epsilon)=\eta \frac{1}{e^{\beta \mu} + e^{\beta \epsilon}},
\end{displaymath}
where $\eta$ is a normalization constant and $\mu$ is the chemical
potential which parameterizes the distribution.  With this, it is
straightforward to find the density as a function of potential, but to
get $\Lambda(\chi)$, Poisson's equation must be solved.  For a $\beta
\mu$ value, we numerically determine Lynden-Bell potential and density
distributions.

\begin{figure}
\scalebox{0.5}{
\includegraphics{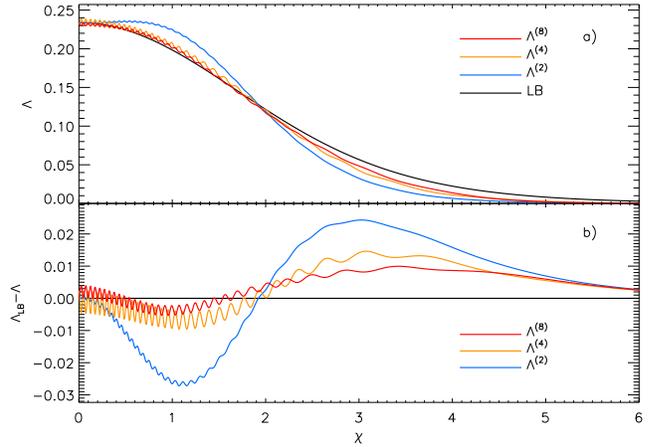}}
\caption{Panel a shows plots of time-independent density distributions
compared to a Lynden-Bell density profile, $\Lambda_{\rm LB}$.  The
Lynden-Bell distribution function has $\beta \mu=1$.  The $A^{(2)}$
time-independent solution has $\Delta=0.5$, the $A^{(4)}$ solution has
$\Delta_2=0.6$ and $\Delta_4=0.15$, and the $A^{(8)}$ solution has
$\Delta_2=0.625$, $\Delta_4=0.15$, $\Delta_6=-0.1$, and
$\Delta_8=-0.08$.  The various models are differentiated in the
legend.  All time-independent solutions have $n_{\rm max}=128$.  Panel
b highlights the differences between the Lynden-Bell and various
time-independent densities. While the agreement is not perfect, it is
clear that higher-order time-independent solutions can produce
densities quite similar to Lynden-Bell models.
\label{lbdcomp}}
\end{figure}

We have decomposed the difference between a normalized Lynden-Bell
distribution function and the equilibrium distribution function into
coefficients.  By removing equilibrium, we focus on just the
perturbation that the Lynden-Bell function represents.  We find that
the $n=0$ coefficients decrease in magnitude with increasing $m$.  As
a result, we expect that higher-order time-independent functions with
$\Delta_m$ values that decrease as $m$ increases should be able to
provide more accurate matches to a Lynden-Bell distribution function.
Without doing an optimized search, we have found a decent
approximation for a Lynden-Bell density, with $\beta \mu=1$, in an
$A^{(4)}$ solution with $\Delta_2=0.6$ and $\Delta_4=0.15$.
Figure~\ref{lbdcomp} shows comparisons between the $A^{(2)}$
($\Delta=0.5$), the $A^{(4)}$ mentioned above, a higher-order
$A^{(8)}$ solution, and the Lynden-Bell densities.  The progression
in the figure supports the conjecture that Lynden-Bell equilibria are
examples of these time-independent solutions.

\subsection{Family Boundaries}\label{bound}

The $A^{(2)}$ family of solutions extends from $\Delta=-1/\sqrt{2}$
to $\Delta \gg 1$.  The lower limit is set by the point at which the
kinetic energy of the system is zero.  As pointed out in \citet{br14},
the kinetic energy is simply related to the $A_{2,0}$ coefficient,
\begin{displaymath}
\beta K = \frac{1}{2} + \sqrt{\sqrt{\pi}} A_{2,0}.
\end{displaymath}
The upper limit to $\Delta$ is undetermined from our explorations.  As
$\Delta$ is increased from 1 to 2, we find that the minimum density
(which occurs at $\chi=0$) appears to slowly, possibly exponentially,
approach zero.

\begin{figure}
\scalebox{0.5}{
\includegraphics{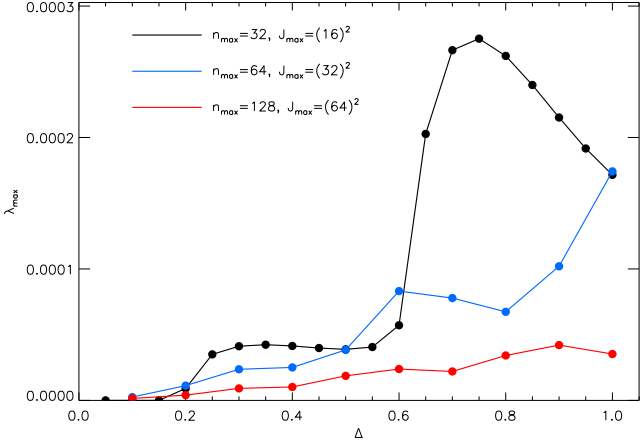}}
\caption{Boundaries of the $A^{(4)}$ family parameters $\Delta_2$ and
$\Delta_4$ that produce density distributions that are positive for
all $\chi$.  The dots represent results from specific $A^{(4)}$
members that have been investigated.  The thick solid lines are linear
fits to the two sets of points.  With $\Delta_2 \la -0.3$, no
continuously-positive-density family member can be created.
\label{ati4bnd}}
\end{figure}

As the $A^{(4)}$ family has two parameters, we have investigated how
$A_{2,0}^{(4)}$ and $A_{4,0}^{(4)}$ interact.  To isolate the impact
of the $A_{4,0}^{(4)}$ value, we initially set $A_{2,0}^{(4)}$ to
zero.  In this situation, any negative value of $A_{4,0}^{(4)}$
results in negative density values.  Similarly, $\Delta_4 \ga 0.7$
also produces negative densities.  As $\Delta_2$ is changed from zero,
the upper limiting value of $\Delta_4$ also changes.
Figure~\ref{ati4bnd} shows the approximate boundary for non-zero
density distributions for the $A^{(4)}$ family.  The points mark
locations where a rough grid search of parameter space result in
negative densities.  The thick, solid lines result from linear fits to
the two sets of dots, while the hatched area denotes $\Delta_2$ and
$\Delta_4$ values that produce positive densities at any location.
The kinetic energy still only depends on the $A_{2,0}^{(4)}$ term, so
there remains the same limit of $\Delta_2=-1/\sqrt{2}$ for positive
kinetic energy.  However, we have not found any system with
continuously positive density (for all $\chi$) when $\Delta_2 \la
-0.3$.  As mentioned previously, numerical simulations can settle down
into states that are nearly, but not quite, Lynden-Bell equilibria.
From the point of view of this work, this behavior reflects that there
is a parameter sub-space in Figure~\ref{ati4bnd} that not only
guarantees non-negative densities, but also produces systems similar
to Lynden-Bell models ($\Delta_4 < \Delta_2$).

\subsection{Stability Analysis}

In addition to identifying these equilibria, we have also investigated
their stability.  Rather than producing a linearized collisionless
Boltzmann equation about the $f_0$ equilibrium, we have linearized
about an non-linear steady state.  This produces a recursion
relation for the coefficients of these linear perturbations $d_{m,n}$.
It is very similar in structure to Equation~\ref{cbeap},
\begin{eqnarray}\label{cbelap}
\dot{d}_{m,n} & = & R_{m,n}^{m-1,n-1} \, d_{m-1,n-1}
  + R_{m,n}^{m-1,n+1} \, d_{m-1,n+1}\nonumber\\ & + & R_{m,n}^{m+1,n-1}
  \, d_{m+1,n-1}  + R_{m,n}^{m+1,n+1} \, d_{m+1,n+1} \nonumber\\
 & - & V_1 + V_2.
\end{eqnarray}
The $R$ matrix elements maintain the same form as before, but in this
case the $V_1$ and $V_2$ terms remain linear in the unknown $d_{m,n}$
coefficients, 
\begin{eqnarray}
\lefteqn{V_1= 2\sqrt{\sqrt{\pi}m(2n+1)} \times}
\nonumber \\
& & \sum_{i \ge 1}^{\infty}
\frac{A_{0,i}}{\sqrt{2i+1}}\sum_{s=0 \atop {\rm even}}^{n+i+1}
\frac{d_{m-1,n+i+1-s}}{2(n+i+1-s)+1} Q_s^{(n,i+1)},
\end{eqnarray}
and
\begin{eqnarray}
\lefteqn{V_2= 2\sqrt{\sqrt{\pi}m(2n+1)} \times}
\nonumber \\
& & \sum_{i \ge 1}^{\infty}
\frac{d_{0,i}}{\sqrt{2i+1}}\sum_{s=0 \atop {\rm even}}^{n+i-1}
\frac{A_{m-1,n+i-1-s}}{2(n+i-1-s)+1} Q_s^{(n,i-1)}.
\end{eqnarray}
Similar to the tactic described in Section~\ref{lprrp}, a matrix
equation can be developed for these recursion relations.  The
eigenvalues of the matrix that represents the right-hand side of
Equation~\ref{cbelap} then provide information about the stability of
the equilibrium described by the $A_{m,n}$.

\begin{figure}
\scalebox{0.5}{
\includegraphics{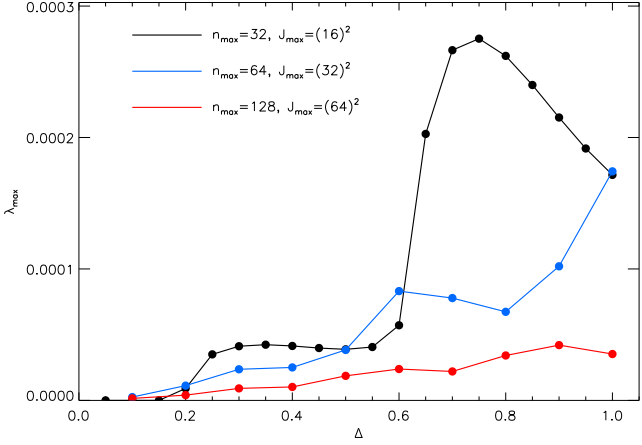}}
\caption{Maximum real eigenvalues for perturbations from a set of
$A^{(2)}$ equilibria.  The three curves represent
results based on $n_{\rm max}=32$, 64, and 128, respectively.  The
points show the eigenvalues as functions of perturbation strength.
Unsurprisingly, larger perturbations tend to result in more unstable
systems.  For these curves, we have linked the size of the
perturbation coefficient system to the values of $n_{\rm max}$.
However, the curve behaviors depend only on $J_{\rm max}$.  For
example, the $n_{\rm max}=128$ curve would look like the $n_{\rm
max}=64$ curve if $J_{\rm max}=(32)^2$ were used instead.
\label{ati2eig}}
\end{figure}

We have tracked the maximum real eigenvalue $\lambda_{\rm max}$ for
several members of the $A^{(2)}$ family, $0.1 \le \Delta \le 1.0$ with
$n_{\rm max}=32$, 64, and 128.  Figure~\ref{ati2eig} shows the general
increase in $\lambda_{\rm max}$ with increasing $\Delta$ across the
$n_{\rm max}$ values.  For this figure, we have set the size of the
perturbation coefficient system based on $n_{\rm max}$.  The
$\mathbfss{R+V}$ matrix that represents the non-linear recursion
relations is square, with optimal dimension $J_{\rm max}=(n_{\rm
max}/2)^2$.  By restricting the size of the $\mathbfss{R+V}$
matrix for a fixed $n_{\rm max}$, we have found that the differences
in the Figure~\ref{ati2eig} curves is due to the size of the
$\mathbfss{R+V}$ matrix.  For example, using an $n_{\rm max}=64$
solution with a $\mathbfss{R+V}$ matrix with size $(16)^2$, not
$(32)^2$, produces a stability curve that looks the same as the
$n_{\rm max}=32$ curve in Figure~\ref{ati2eig}.  As a result, it
appears that in the large $J_{\rm max}$ (and $n_{\rm max}$) limit,
these solutions become more and more stable.

\begin{figure}
\scalebox{0.5}{
\includegraphics{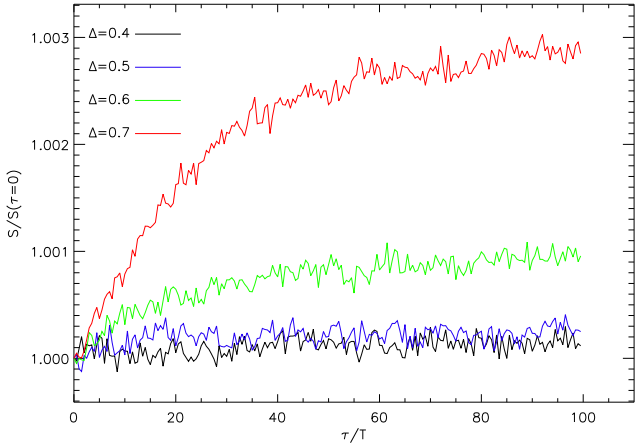}}
\caption{Entropy evolutions for several $N$-body ensembles
representing various $A^{(2)}$ family members with $n_{\rm max}=64$.
The growth of instability with rising $\Delta$ values is evident.  The
stark difference between $\Delta<0.5$ and $\Delta>0.5$ suggests that
$N$-body effects are masking the smoother expected rise seen for the
$n_{\rm max}=64$ curve in Figure~\ref{ati2eig}.  The $N$-body systems
have a lower, effective $J_{\rm max}$.
\label{nba2ent}}
\end{figure}

\begin{figure}
\scalebox{0.5}{
\includegraphics{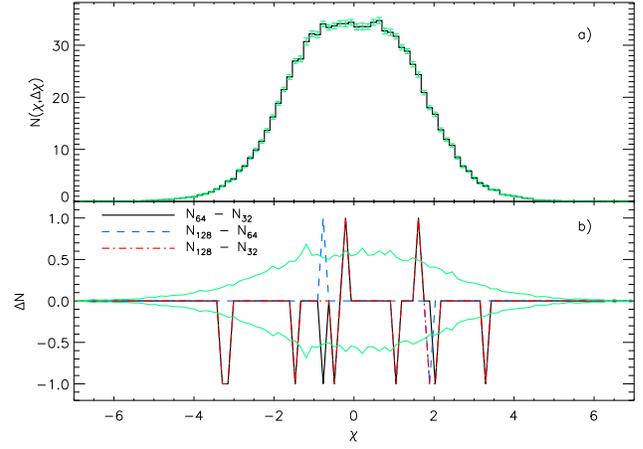}}
\caption{Panel a shows the ensemble average spatial distribution of
$N$-body particles ($N=1024$) when an $A^{(2)}$ solution with
$\Delta=0.5$ and $n_{\rm max}=128$ is used as the initial condition.
Error bars reflect the error-in-the-mean value for each bin.
Comparisons with results based on other $n_{\rm max}$ solutions is
summarized in panel b.  The different line styles show differences
between the three $n_{\rm max}$ values investigated; for example,
changes between $n_{\rm max}=64$ and $n_{\rm max}=32$ $N(\chi,\Delta
\chi)$ distributions (denoted by $N_{64}-N_{32}$ in the panel).  The
error-in-the-mean values are indicated by the thin lines.  Maximum
differences of 1 particle per bin suggest finite $N$ effects are
masking any influence that the smoothness of solutions with larger
$n_{\rm max}$ values may impart to stability.
\label{nbiccomp}}
\end{figure}

\begin{figure}
\scalebox{0.5}{
\includegraphics{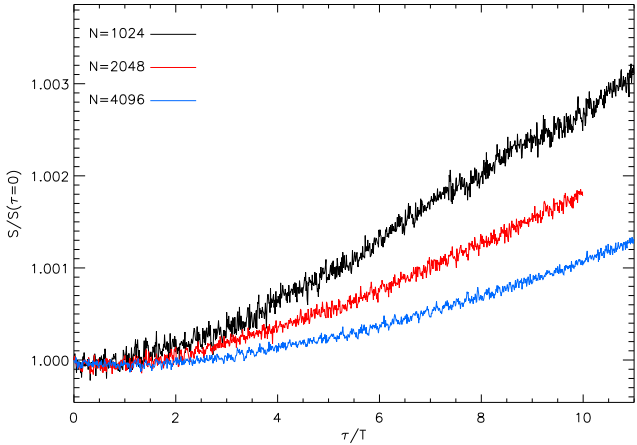}}
\caption{Plots of $N$-body entropy evolutions for ensembles with the
same $A^{(2)}$ solution initial conditions with $\Delta=1.0$ and
$n_{\rm max}=128$.  The ensembles differ only in their particle
numbers, as indicated in the legend.  The $N=2048$ ensemble evolutions
extend only to $\tau=10$ while only the initial stages of the other
evolutions are shown.  The trend for slower rise in entropy given
larger particle numbers indicates that the time-independent solution,
with less instability, is reproduced more accurately.
\label{nbsnp}}
\end{figure}

These results have been compared with $N$-body simulations like those
described earlier.  Initial conditions are drawn from $A^{(2)}$
solutions, and the systems are allowed to evolve for 100 $T$.
Figure~\ref{nba2ent} shows the entropy evolutions of several $N$-body
ensembles based on $n_{\rm max}=64$ solutions.  The $N$-body curves
shown represent time-independent solutions with $\Delta$ values around
0.5.  The dichotomy of $N$-body behaviors about $\Delta=0.5$ is more
reminiscent of the behavior seen for the $J_{\rm max}=(16)^2$ curve in
Figure~\ref{ati2eig}.  We suggest that the impact of finite particle
numbers degrades the equivalent modal resolution.  A point in support
of this is the fact that initial $N$-body density distributions based
on $n_{\rm max}=32$, 64, and 128 solutions are essentially
indistinguishable.  Figure~\ref{nbiccomp}a shows the ensemble-average
initial spatial distribution of particles when an $A^{(2)}$ solution
with $\Delta=0.5$ and $n_{\rm max}=128$ is used.
Figure~\ref{nbiccomp}b shows differences in the number of particles
per bin when $n_{\rm max}=32$ or $n_{\rm max}=64$ solutions are used.
To focus on variations due only to $n_{\rm max}$, the random number
sequence used for each ensemble has been held fixed.  The thin lines
in Figure~\ref{nbiccomp} indicate the error in the mean for the
ensemble average at each bin (which are also the error bars in
Figure~\ref{nbiccomp}a).  Essentially, our $N$-body simulations cannot
resolve the underlying distribution function well enough to allow the
stability differences to appear.  Further support of this idea is
given in Figure~\ref{nbsnp}.  For an $A^{(2)}$ solution with
$\Delta=1.0$ and $n_{\rm max}=128$, we have run additional $N$-body
ensembles with $N=2048$ and $N=4096$.  The slower rise in entropy with
larger particle number is in line with our hypothesis that more
particles result in a higher fidelity time-independent solution
representation, which results in less instability.  We have also
created a $N=2048$ ensemble with $n_{\rm max}=64$ to confirm that this
behavior occurs generally.

The stability analysis can be extended to members of the $A^{(4)}$
family as well.  While we have not done an exhaustive search over the
$(\Delta_2,\Delta_4)$ plane, we find that systems that would lie within
the hatched region of Figure~\ref{ati4bnd} have maximum positive
eigenvalues that are approximately an order of magnitude smaller than
$\lambda_{\rm max}$ for a system that lies outside the hatched region.
Specifically, the Lynden-Bell-like system with $\Delta_2=0.6$ and
$\Delta_4=0.15$ and a mildly perturbed system with
$\Delta_2=\Delta_4=0.2$ have $\lambda_{\rm max} \approx 1 \times
10^{-4}$ and produce stable $N$-body systems.  On the other hand, the
$\Delta_2=0.2$ and $\Delta_4=-0.4$ model is not stable in an $N$-body
simulation and has $\lambda_{\rm max} \approx 1 \times 10^{-3}$.
From the values in Figure~\ref{ati2eig}, a factor of 10 indicates a
significant difference in stability.

Non-linear solutions are approximated up to some $n_{\rm max}$.
Our results indicate that those solutions are stable in the limit
$n_{\rm max} \rightarrow \infty$.  $N$-body realizations of these
non-linear steady states appear to have a perturbation amplitude
dependence (Figure~\ref{nba2ent}), but we argue that this is a result
of finite particle numbers.  Modest numbers of particles do not allow
simulated systems to capture the small, but important, differences
between models with small and large $n_{\rm max}$.  Simulations with
increasing $N$ result in a clear trend towards increasing stability.

\section{Summary}

We have presented a procedure for calculating time-independent
solutions of arbitrary perturbations of one-dimensional gravitating
systems.  Both test-particle and self-gravitating systems can be
analyzed with this approach.  Sets of coefficients that describe
Hermite-Legendre polynomial products form the time-independent
solutions.  In the case of linear perturbations from equilibrium, the
solutions are independent modes.  Suites of highly efficient and
accurate $N$-body simulations have been created to test predictions
based on steady-state solutions.

Starting with linear perturbations, we find that there are two routes
to determining these coefficient sets.  For what we term $E$ modes,
coefficient sets are limited in their Hermite index.  These modes are
directly related to energies of a system.  As expected with time
independence, all $E$ modes are in virial equilibrium.  The first $E$
mode has non-zero energy and represents changing the temperature of
the separable equilibrium state.  The alternative $B$ modes are formed
by limiting the Legendre index, but the lack of an obvious, related
physical quantity makes them less appealing than their $E$-mode
analogues.  However, steady states can be predicted just as well using
either set of modes.

For large amplitude perturbations, we follow only the $E$ mode
approach.  Non-linear terms in the collisionless Boltzmann equation
necessitate an iterative solution approach to solving the coupled
coefficient recursion relations.  There are boundaries on the
parameters of these solutions based on maintaining positive kinetic
energy and continuously positive density values.  We have found that a
subset of these solutions are similar to Lynden-Bell models.
Analyzing Lynden-Bell distribution functions suggests that increasing
the maximum Hermite order of solutions allows for better
approximations to the Lynden-Bell form.  Analyzing the stability of
non-linear perturbation solutions via coefficient dynamics and
$N$-body simulations indicates that physically relevant (positive
kinetic energy and density) steady states are stable.  However, that
stability can be upset by insufficient particle numbers in $N$-body
simulations.

Arguably the most important use of these time-independent modes is the
prediction of steady states from linear perturbation initial
conditions.  We find that for modest strength perturbations in
self-gravitating systems, any non-linearities present are of the same
order as statistical uncertainties and that the time-independent modes
accurately predict the simulated steady states.  Unfortunately, such
an approach is not possible for non-linear initial conditions.  At
best, one might be able to use coefficient values from initial
conditions to determine what families of time-independent solutions
may be present.

\end{document}